\documentstyle[12pt]{article}
\newcommand{\eq}{\begin{equation}}
\newcommand{\en}{\end{equation}}
\newcommand{\eqn}{\begin{eqnarray}}
\newcommand{\enn}{\end{eqnarray}}

\begin{document}
\begin{titlepage}
\begin{flushright}
THEP-93-4 \\
Feb. 1993
\end{flushright}
\vspace{2cm}
\begin{center}
{\LARGE
Gamma Ray Bursts and CETI}

\vspace{1cm}
{\large Frank D. (Tony) Smith, Jr.} \\
Department of Physics \\
Georgia Institute of Technology \\
Atlanta, Georgia 30332 \\
\vspace{1cm}
{\bf Abstract}
\end{center}
Gamma ray burst sources are isotropically
distributed.  They could be located at distances
$\sim 1000$ AU. (Katz \cite{JK92}) \\
GRB signals have many narrow peaks that are
unresolved at the millisecond time resolution
of existing observations. \cite{JK87} \\
CETI could use stars as gravitational lenses
for interstellar gamma ray laser beam communication. \\
Much better time resolution of GRB signals
could rule out (or confirm?) the speculative hypothesis
that GRB = CETI.

\vspace{1cm}
\normalsize
\footnoterule
\noindent
{\footnotesize (\copyright 1993
Frank D. (Tony) Smith, Jr., 341 Blanton Road,
Atlanta, Georgia 30342 USA. \\
P. O. Box for snail-mail:  P. O. Box 430,
Cartersville, Georgia 30120 USA \\
e-mail: gt0109e@prism.gatech.edu and
fsmith@pinet.aip.org }
\end{titlepage}
\setcounter{footnote}{0}
\section{Introduction}
\setcounter{equation}{0}

Gamma ray burst (GRB) sources were discovered
by U. S. and Russian satellites searching for
nuclear explosions. \cite{JK87,KL}

Observations from the Compton Observatory
made GRBs even more enigmatic, because they
were found to be isotropically distributed about
the earth.  \cite{PT}

The isotropic distribution ruled out sources
confined to the galactic plane, leaving three
possibilities:

1.  GRBs are at cosmological distances;

2.  GRBs are at galactic halo distances; and

3.  GRBs are relatively local, in the solar
neighborhood.

\vspace{12 pt}

In this speculative paper, it is assumed that
GRBs are in the solar neighborhood at
a distance $\sim 1000$ AU, roughly the distance
of an inner Oort cloud.

\vspace{12pt}

How energetic must Oort cloud GRBs be?
GRBs located 100 light-hours (720 AU)
from the sun radiate about $10^{26}$ erg,
assuming that they are spherically symmetric
sources of gamma rays.  \cite{PT}

\vspace{12pt}

Oort cloud GRBs have been discussed by Katz
\cite{JK92}, a paper entitled "{\it A Burst
of Speculation}".

As an energy source, Katz proposes (in a
section entitled "{\it A Crazy Idea}") cometary
collisions.
The idea of colliding cometary ice producing
gamma rays really may not be so crazy,
particularly taking into account such poorly
understood phenomena as sonoluminescence
in water due to cavitation (Maddox \cite{JM}
says that some researchers argue that
the cavitation bubble could produce
plasmas with black-body temperature
at least $\sim 100,000$ K.
(For comparison, 1 MeV $\approx10^{10}$ K)).

\vspace{12pt}

Although Katz may be right about Oort
cloud comets, there are other possibilities,
including CETI (Communications from
ExtraTerrestrial Intelligences).

"It may be that the very first sign we ever
find is some weird phenomenon that a
conventional astronomer ses who wasn't
even thinking about SETI at that time.", a
quote from Thomas McDonough \cite{ST},
might apply to GRBs.

\vspace{12pt}

The purpose of this paper is to discuss that
possibility, and to propose that future
observations (particularly better time
resolution of GRB signals) should be
undertaken to rule out or confirm the
CETI possibility.

\section{CETI}

CETI (Communications from ExtraTerrestrial
Intelligences) should be distinguished from
SETI (Search for (or Signals from)
ExtraTerrestrial Intelligences).

\vspace{12pt}

SETI (the subject of the first Thomas McDonough
quote \cite{ST}) assumes that the ETs have
set up a beacon to attract our attention.

\vspace{12pt}

CETI assumes that the ETs don't really care
whether or not they attract our attention, but
are busily communicating among themselves
and that we might be able to eavesdrop on their
conversations.

\vspace{12pt}

Beacons should be bright signals without
much structure.  Conversations should have
a lot of structure.

\vspace{12pt}

Since GRB signals fluctuate erratically during
their duration ($\sim 30$ sec), and have many narrow
peaks whose width is unresolved by existing
observations ($\sim$ milliseconds) \cite{JK87,PT},
it seems that, if GRBs are ET,
they are more likely to be CETI than SETI.

\vspace{12pt}

Therefore, this paper discusses the CETI possibility.

\vspace{12pt}

\section{Stars as Gravity Lenses}

If ETs wanted to set up a galactic version of Internet,
how would they do it?

\vspace{12pt}

Stars act as gravitational lenses.

The focal length of the sun as a gravitational lens
is about 540 AU.  Any beam of electromagnetic
radiation (whether light, radio waves, or gamma
rays) hitting the sun from $\approx 540$ AU or farther
out is focussed by the sun's gravitational field into
a beam that could be used for communication.

\vspace{12pt}

"If we could  but see it, this is what every star in
the galaxy looks like, sort of a sea urchin if you will,
a star making images of every other star, starting
at the minimum distance and going out to infinity.
The casting of very high-resolution images of the
whole universe on the sky - and these images are in
focus at all distances - is a really remarkable thing."
(Drake discussing the ideas of Eshleman in the book
of McDonough \cite{TM})

\vspace{12pt}

"Each star produces hundreds of billions of tubes of
light, one for every other star, so there are hundreds
of billions of hundreds of billions of tubes of light
in our galaxy." (McDonough \cite{TM})

\vspace{12pt}

The network of tubes of light would be a natural
foundation for the ETs to build on to construct their
galactic CETI Internet.

\vspace{12pt}

For ET to set up a transmitter/receiver station
using the sun as a gravity lens, the ET would have
to put it at least $\approx 540$ AU away from the sun,
but they would want it as near as practical to minimize
the time and energy spent in maneuvering from
one target to another.

Therefore, I assume that the ET CETI stations using
the sun are \linebreak
$\sim 500-1000$ AU from the sun, and hence
$\sim 500-1000$ AU from the earth.

\vspace{12pt}

The tracking and maneuver requirments of using
the Oort cloud neighborhood of the sun as a
transmitter/receiver station in a galactic Internet
involves a technology beyond my ability to outline,
but I presume that it is within ET capability.

\vspace{12pt}

To maximize the information their signals could carry,
the ET would use coherent signals with short wavelengths:
gamma ray lasers.

\vspace{12pt}

Creation, modulation and demodulation of such a
gamma ray signal involves a technology beyond
my ability to outline, but I presume that it is
within ET capability.

Even we can now use lasers for free-space laser
communication \cite{SPIE}, and the search for ET laser signals
has been proposed by Townes and Betz (infrared) and
Kingsley (optical) \cite{ST,SPIE}.

\vspace{12pt}

If GRBs are the gamma ray CETI signals from such a
galactic Internet, then the duration of each message
is $\sim 30$ sec, and there are about 2 messages/day.
\cite{PT}

\vspace{12pt}

What are the energy requirements?

If they were spherically symmetric sources,
GRBs located 100 light-hours (720 AU)
from the sun would radiate about $10^{26}$ erg. \cite{PT}

However, if the ET were beaming the gamma rays in
the direciton of the sun, confining the beam to
$\sim 10^{-4}$ steradian, the earth would still be
in the beam and the required energy would be
$\sim {10^{26}  \times 10^{-4}} / {4 \pi}$
$\approx 10^{21}$ erg.

For comparison, $\sim 10^{21}$ erg is roughly the energy of
fission of 1 kg of $U^{235}$, or of the annihilation of
1 gm of matter by antimatter.

\vspace{12pt}

How can we tell whether or not the GRBs we see are
signals on a galactic gamma-ray laser CETI Internet?

\section{Need for Better Time Resolution}

To tell whether or not the GRBs we see are
signals on a galactic gamma-ray laser CETI Internet,
we need much better time resolution of the GRB
signals than $\sim$ milliseconds.

On that scale, there is a lot of substructure \cite{PT},
and we have not yet resolved the time structure of the
GRB signals \cite{JK87}.

\vspace{12pt}

One last quote from McDonough \cite{TM}:

"Perhaps the first signs of an extraterrestrial
civilization are already sitting on magnetic tape,
deep in the bowels of the U. S. National Security
Agency or the Soviet KGB."

The NSA may not yet have GRB signals with time
resolution much better than milliseconds.

However, if GRB = CETI, the skills of the NSA will
probably be required to figure out what the ETs are
saying to each other.



\begin{thebibliography}{99}
\bibitem{JK87} J. Katz, ``High Energy Astrophysics",
Addison-Wesley (1987).
\bibitem{KL} R. Klebesadel, I. Strong, and R. Olson,
{\it Ap. J. (Lett.)} {\bf 182} L85 (1973).
\bibitem{PT} B. Schwarzchild, {\it Compton
Observatory Data Deepen the Gamma Ray
Burster Mystery}, {\it Physics Today}
(February 1992), 21.
\bibitem{JK92} J. Katz, {\it A Burst of Speculation}
ASTRO-PH/9211001 (1992).
\bibitem{JM} J. Maddox, {\it Sonoluminescence in
from the dark}, {\it Nature} {\bf 361} (4 Feb 1993) 397.
\bibitem{TM} T. McDonough, ``The Search for
Extraterrestrial Intellingence", Wiley (1987).
\bibitem{ST} R. Naeye, {\it SETI at the
Crossroads}, {\it Sky and Telescope}
(November 1992), 507.
\bibitem{SPIE} OE/LASE 93, {\it Free-Space Laser
Communication Sessions}, SPIE Los Angeles,
January 1993.
\end{thebibliography}
\end{document}